\documentclass[12pt,a4paper,twoside]{article}
\usepackage[T1]{fontenc}
\usepackage{amsmath,amsfonts,amssymb,amsthm,mathabx}
\usepackage{cite}
\usepackage{hyperref}
\usepackage{graphicx}
\usepackage{pdfsync}
\newcommand{\be}{\begin{equation}}
\newcommand{\ee}{\end{equation}}
\newcommand{\bea}{\begin{eqnarray}}
\newcommand{\eea}{\end{eqnarray}}

\newcommand{\der}{\partial}

\def\d_Vphi{\mathrm{d}_V\hspace{-0.06em}\phi}
\def\d_Vphibar{\mathrm{d}_V\hspace{-0.06em}\bar\phi}
\def\d_Vxi{\mathrm{d}_V\hspace{-0.06em}\xi}

\def\be{\begin{eqnarray}}
\def\ee{\end{eqnarray}}
\def\beann{\begin{eqnarray*}}
\def\eeann{\end{eqnarray*}}
\def\beq{\begin{equation}}
\def\eeq{\end{equation}}
\def\ba{\begin{array}}
\def\ea{\end{array}}
\def\ben{\begin{enumerate}}
\def\een{\end{enumerate}}
\def\bea{\begin{eqnarray}}
\def\eea{\end{eqnarray}}

\def\5{\bar }
\def\6{\partial }
\def\7{\hat }
\def\4{\tilde }
\advance\textheight\topskip
\renewcommand{\tilde}{\widetilde}
\renewcommand{\hat}{\widehat}

\renewcommand{\geq}{\,{\geqslant}\,}

\newcommand{\binner}[2]{%
  {\langle}\kern-4.15pt{\langle}#1{,}\,#2{\rangle}\kern-4.15pt{\rangle}}

\newcommand{\half}{\frac{1}{2}}

\newcommand{\ffrac}[2]{\raisebox{.5pt}%
  {\footnotesize$\displaystyle\frac{#1}{#2}$}\kern1pt}

\def\cD{\mathcal{D}}

\DeclareFontFamily{OT1}{rsfs}{} \DeclareFontShape{OT1}{rsfs}{m}{n}{
<-7> rsfs5 <7-10> rsfs7 <10-> rsfs10}{}
\DeclareMathAlphabet{\mycal}{OT1}{rsfs}{m}{n}

\begin{document}

\title{One loop partition function of three-dimensional
   flat gravity}

\author{Glenn Barnich, Hern\'an Gonz\'alez, Alex Maloney, Blagoje Oblak}

\date{}

\def\mytitle{One-loop partition function of three-dimensional
   flat gravity}

 \pagestyle{myheadings} \markboth{\textsc{\small G.~Barnich,
     H.~A.~Gonz\'alez, A.~Maloney, B.~Oblak}}{%
   \textsc{\small 1-loop partition function for 3d asymptotically flat gravity}}

\addtolength{\headsep}{4pt}

\begin{centering}

  \vspace{1cm}

  \textbf{\Large{\mytitle}}

  \vspace{1.5cm}

  {\large G.~Barnich$^a$, H.~A.~Gonz\'alez$^a$, A.~Maloney$^b$,
    B.~Oblak$^a$}

\vspace{.5cm}

\begin{minipage}{.9\textwidth}\small \it  \begin{center}
   $^a$ Physique Th\'eorique et Math\'ematique \\ Universit\'e Libre de
   Bruxelles and International Solvay Institutes \\ Campus
   Plaine C.P. 231, B-1050 Bruxelles, Belgium
 \end{center}
\end{minipage}

\vspace{.5cm}

\begin{minipage}{.9\textwidth}\small \it  \begin{center}
   $^b$ McGill Physics Department, 3600 rue University, \\
   Montr\'eal, QC H3A 2T8, Canada
 \end{center}
\end{minipage}

\end{centering}

\vspace{1cm}

\begin{center}
  \begin{minipage}{.9\textwidth}
    \textsc{Abstract}. In this note we point out that the one-loop
    partition function of three-dimensional flat gravity, computed
    along the lines originally developed for the anti-de Sitter case,
    reproduces characters of the BMS$_3$ group.
  \end{minipage}
\end{center}

\vspace{1cm}
\thispagestyle{empty}
\newpage

% \begin{small}
% {\addtolength{\parskip}{-2pt}
%  \tableofcontents}
% \end{small}
% \thispagestyle{empty}
% \newpage

\section{Introduction}
\label{sec:introduction}

Three-dimensional gravity has emerged as an important testing ground for our ideas about quantum gravity.  
Following earlier work \cite{Witten:2007kt, Maloney:2007ud}, it was shown that at one loop
the partition function of AdS$_3$ gravity
 is a character of the Virasoro algebra \cite{Giombi:2008vd} (see \cite{David:2009xg} for an improved
derivation and generalizations).  This is in line
with the analysis of the asymptotic symmetry group of AdS$_3$ gravity by Brown and Henneaux
\cite{Brown:1986nw}, and is a direct check of the fact that, at the quantum mechanical level, the bulk gravity states organize into representations of the conformal group. 
In this note we point out that
a similar one-loop computation in the technically simpler case of
flat gravity reproduces the vacuum character of the BMS$_3$ group. The
latter is the symmetry group of asymptotically flat gravity at null
infinity \cite{Ashtekar1997,Barnich:2006avcorr}, whose characters have
been recently worked out in \cite{Oblak:2015sea} (see also
\cite{Barnich:2014kra,Barnich:2015uva} for details on BMS$_3$
representations).

\section{The set-up}
\label{sec:set-up}

We wish to study the gravitational
partition function
\begin{equation}
  \label{eq:1}
  Z[\beta,\theta]=\int \cD g\ e^{-\frac{1}{\hbar}S_E},\quad
  S_E=-\frac{1}{16\pi G}\int d^3x\, \sqrt g R+B. 
\end{equation}
Here $S_E$ is the Euclidean Einstein action (with no cosmological constant), $B$ is an
appropriate boundary term and $\beta,\theta$ are the inverse
temperature and angular potential. 
The parameters $\beta, \theta$ determine a quotient
$\mathbb R^3/{\mathbb Z}$ of flat Euclidean space, representing
``thermal spinning $\mathbb R^3$'':
\begin{equation}
  \label{eq:2}
  \begin{split}
  ds^2&=dx_1^2+dx_2^2+dx_3^2=dy^2+d\rho^2+\rho^2d\varphi^2,
  \\
&  \quad (y,\varphi)\sim 
\gamma(y,\varphi)=(y+\beta,\varphi+\theta) 
\end{split}
\end{equation}
where ${\mathbb Z}$ is the discrete group generated by the identification $\gamma$.

We wish to compute $Z[\beta, \theta]$ in a perturbative expansion around the background \eqref{eq:2}.
The partition function \eqref{eq:1}
can be expanded as 
\begin{equation}
  \label{eq:3}
  \hbar \ln Z[\beta,\theta]=- S^{(0)} + \hbar S^{(1)} + \hbar^2 S^{(2)} + \dots
  \end{equation}
where $S^{(0)}$ is the classical action and $S^{(i)}$ an $i-$loop correction. We will focus on $S^{(0)}$ and 
$S^{(1)}$.

There are several ways to compute  $S^{(0)}$ from the
classical action, including boundary terms. The most direct is to use the
Hamiltonian form of the action. In this case, the surface term at
infinity is, by definition, determined by the surface charges times
their chemical potentials \cite{Regge:1974zd,Banados:1992wn}. For flat
space, this is the only contribution and the on-shell action is
automatically finite. Note however that the overall normalization of
the charges is not fixed. In other words, requiring a well defined
variational principe and a finite on-shell action for a class of
spacetimes does not fix the ambiguity that consists in adding a finite
combination of the quantities that are held fixed in the variational
principle. An additional criterion is needed to lift this ambiguity.

In order to make the classical part of the partition function
invariant under the analog of modular $S$-transformations in the flat
case \cite{Barnich:2012xq,Bagchi:2012xr}
\begin{equation}
  \beta\to \frac{4\pi^2\beta}{\theta^2},\quad \theta\to
  -\frac{4\pi^2}{\theta}\label{eq:4},  
\end{equation}
the charges have to be normalised with respect to the null orbifold
\cite{Horowitz:1990ap}, $ds^2=-2dudr+r^2d\phi^2$. This puts mass and
angular momentum of flat space at $M=-\frac{1}{8G}, J=0$
\cite{Barnich:2006avcorr}, and the tree-level Euclidean action for flat space is
\begin{equation}
S^{(0)}=-\frac{\beta}{8G},\label{eq:15}
\end{equation}
a result originally derived along slightly different lines in
\cite{Bagchi:2013lma}. For cosmological solutions
\cite{Ezawa:1992nk,Cornalba:2002fi,Cornalba:2003kd,Barnich:2012aw}
with metric
\begin{equation}
ds^2= 8GM du^2-2dudr + 8GJdud\phi +r^2d\phi^2%\label{eq:8}
\nonumber
\end{equation}
and $M\geq 0$, the Euclidean action takes the value 
\begin{equation}
  %\label{eq:9}
  S^{(0)}=-\frac{\pi^2\beta}{2G\theta^2}, 
  \nonumber
\end{equation}
which can be obtained from the flat space result (\ref{eq:15}) through the
transformation \eqref{eq:4}. 

One can then use the fact that three-dimensional gravity has no local degrees of freedom to understand the nature 
of the quantum corrections $S^{(i)}$ to this classical action, following \cite{Maloney:2007ud}.  One might 
naively think that -- as there are no local degrees of freedom to run in loops -- all  loop corrections 
must vanish.  This is not quite the case, however, since there may not be complete cancellation between the 
graviton loops and the ghost loops which arise from gauge fixing.  In theories with topological degrees of 
freedom, such as Chern-Simons theory, this is typically the case.  Given the similarities between three-dimensional gravity and 
Chern-Simons theory we therefore expect that the $S^{(i)}$ may be non-trivial.
  
The perturbative contributions $S^{(i)}$ can in principle be obtained by quantizing a classical phase space, 
which in the present case is the space of metrics which are smoothly connected to the classical solution 
\eqref{eq:2}, modulo an appropriate group of diffeomorphisms. We must impose boundary conditions on the 
metric at asymptotic infinity, following \cite{Ashtekar1997,Barnich:2006avcorr}, which require that we only 
include metrics with finite values of the BMS charges. Only those diffeomorphisms which vanish sufficiently 
quickly at asymptotic infinity -- in the sense that they do not change the values of these charges -- are 
regarded as gauge transformations.  One can then obtain the classical phase space by applying to the metric 
\eqref{eq:2} the diffeomorphisms which do not vanish sufficiently quickly at infinity.  Since there are no 
local degrees of freedom, the $entire$ phase space of solutions which are continuously connected to \eqref{eq:2} is constructed by this group 
of non-trivial diffeomorphisms. More precisely, since the transformation law of asymptotically 
flat metrics under BMS$_3$ is given by the coadjoint representation, the perturbative 
contributions $S^{(i)}$ are obtained by quantizing the coadjoint orbit of flat space under the asymptotic 
symmetry group. The resulting Hilbert space will naturally be a representation of BMS$_3$, and the partition 
function a character of BMS$_3$.  

We therefore expect that the perturbative partition function should be given by the vacuum BMS$_3$ character described in \cite{Oblak:2015sea}.  This result is one loop exact: $S^{(1)}\ne 0$, but $S^{(i)} = 0$ for all $i\ge 2$.
The one-loop exactness of the perturbative gravity partition function was already observed in AdS$_3$ gravity \cite{Maloney:2007ud}.

The above argument, while appealing, has the structure of the BMS group built in from the beginning.  It would be preferable to verify this expectation from a direct gravitational computation, as this would give a check of the BMS structure of flat space gravity at the quantum level.  We will now describe the direct computation of $S^{(1)}$, leaving higher loop corrections for future work.

\section{Evaluation of the determinants}
\label{sec:computation}

We follow closely \cite{Giombi:2008vd}, to which we refer for more
details. The one-loop contribution to the partition function is given by
  \begin{equation}
  S^{(1)}=-\half \ln\det \Delta^{(2)}+\ln\det \Delta^{(1)}-\half
  \ln\det \Delta^{(0)}.
  \label{putTogether}
\end{equation}
Here $\Delta^{(2)},\Delta^{(1)},\Delta^{(0)}$ are the
kinetic operators which arise in the linearized expansion of general relativity around the flat background metric \eqref{eq:2}, including ghosts.
They are the Laplacian operators for a massless, traceless symmetric tensor, a vector, and a scalar,
respectively (see also e.g.~\cite{Christensen:1979iy}).

The determinants are evaluated using the heat kernel approach, which
is straightforward in flat space.  
The heat kernels $K$, $K_{\mu\mu'}$ and $K_{\mu\nu,\mu'\nu'}$ are defined to be solutions to the differential 
equations
\begin{equation}
  %\label{eq:16}
  \begin{split}
   & (\Delta^{(0)}-m^2-\der_t)K(t;x,x')=0, \\
   & (\Delta^{(1)}-m^2-\der_t)K_{\mu \mu'}(t;x,x')= 0, \\
   & (\Delta^{(2)}-m^2-\der_t)K_{\mu\nu,\mu'\nu'}(t;x,x') =0,\\
  \end{split}
  \nonumber
\end{equation}
with boundary conditions at $t=0$:
\begin{equation}
\begin{split}
  K(0;x,x')&=\delta^{(3)}(x-x'), \\
   K_{\mu\nu'}(0;x,x')&=\delta^{(3)}(x-x')\delta_{\mu\mu'}, \\
 K_{\mu\nu,\mu'\nu'}(0;x,x')&=\half(\delta_{\mu\mu'}\delta_{\nu\nu'}
    +\delta_{\mu\nu'}\delta_{\nu\mu'}-\frac{2}{3}\delta_{\mu\nu}
    \delta_{\mu'\nu'})\delta^{(3)}(x-x').
    \end{split}
    \nonumber
\end{equation}
We have introduced a mass parameter $m$, so that these heat kernels encode the spectra of the massive kinetic operators $\Delta^{(i)}-m^2$.
Defining the world function
\begin{equation}
%   \label{eq:12}
  \sigma=\half |x-x'|^2
  \nonumber
\end{equation}
the heat kernels are
\begin{equation}
%   \label{eq:10}
  \begin{split}
& K(t;x,x')=\frac{1}{(4\pi t)^{\frac{3}{2}}}
e^{-m^2t-\frac{\sigma}{2t}}, \\
& K_{\mu \mu'}(t;x,x')=- K(t;x,x') \der_{\mu}
 \der_{\mu'}\sigma,\\
& K_{\mu\nu,\mu'\nu'}(t;x,x')=\half K(t;x,x')
\left(\der_\mu\der_{\mu'}\sigma\der_\nu\der_{\nu'}\sigma+
\der_\mu\der_{\nu'}\sigma\der_\nu\der_{\mu'}\sigma
-\frac{2}{3}\delta_{\mu\nu}\delta_{\mu'\nu'}\right). 
  \end{split}
  \nonumber
\end{equation}

The essential point is that the heat kernels obey linear differential equations, so given the heat kernels in 
flat space one can readily obtain the heat kernel in a quotient of flat space using the method of images.
For thermal spinning $\mathbb R^3$ we have 
\begin{equation}
K^{\mathbb{R}^3/{\mathbb Z}}(t,x,x)
=
\sum_{n\in\mathbb{Z}}
~K(t,\sigma(x,\gamma^n x)).
\nonumber
\end{equation}
The Euclidean distance between a
point and its $n$-th image under $\gamma$ is
\begin{equation}
%   \label{eq:17}
  \sigma(x,\gamma^nx)=\half n^2\beta^2  +2\: {\rm sin}^2\left(\frac{n\theta}{2}\right) 
\rho^2
\nonumber
\end{equation}
so that 
\begin{equation}
%   \label{eq:18}
  K^{\mathbb{R}^3/{\mathbb Z}}(t,x,x)= K(t;x,x)+
\frac{2 e^{-m^2 t} }{(4\pi 
t)^{3/2}} \sum^{\infty}_{n=1} e^{ -{\rm sin}^2\left(\frac{n\theta}{2}\right) 
\frac{\rho^2}{t}-\frac{n^2\beta^2}{4t}}. 
\nonumber
\end{equation}
%Using that $\Gamma(-\frac{3}{2})=\frac{4\sqrt \pi}{3}$, this implies that 

We can now extract the determinants of the operators $\Delta^{(i)} - m^2$ by integrating these
heat kernels:
\begin{equation}
% \label{DeltaZ}
\begin{split}
  -\ln \det \left(\Delta^{(0)}-m^2\right)&=\int^{\infty}_0 \frac{dt}{t}
  \int^{\beta}_0 dy \int^{2\pi}_0 d\varphi \int^{\infty}_0 \rho d\rho \:
  K^{\mathbb{R}^3/{\mathbb Z}}(t,x,x),\\ 
  &=\frac{m^{3}}{6\pi} {\rm
    Vol}(\mathbb{R}^3/{\mathbb Z})+2\sum^{\infty}_{n=1}\int^{\infty}_{0} dt
  \: (2\pi \beta) \frac{e^{-m^2 t}}{(4\pi
    t)^{3/2}} \frac{e^{-\frac{n^2\beta^2}{4t}}}{2
    \sin^2\left(\frac{n\theta}{2}\right)}, \\ 
  &=\frac{m^{3}}{6\pi} {\rm
    Vol}(\mathbb{R}^3/{\mathbb Z})+\sum^{\infty}_{n=1} \frac{2 e^{-n \beta 
      m}}{n |1-e^{in\theta}|^2}
\end{split}
\nonumber
\end{equation}
Defining
$\hat{g}^{\: \mu \nu'}=g^{\mu \rho} \der_{\rho}(\gamma^n x)^{\nu'}$
and using $\hat{g}^{\: \mu \nu'}\der_{\mu}\der_{\nu'}\sigma=-1-2\cos(n\theta)$,
\begin{equation}
% \label{Deltav}
\begin{split}
  -\ln \det\left(\Delta^{(1)}-m^2\right)&=\int^{\infty}_0 \frac{dt}{t}
  \sum_{n\in\mathbb{Z}}\int d^3x \sqrt{g} \ \hat{g}^{\: \mu
    \nu'}K_{\mu \nu'}(t,\sigma(x,\gamma^n x))\\
  &=\frac{m^3}{2\pi}{\rm Vol}({\mathbb
    R^3}/{\mathbb Z})+\sum^{\infty}_{n=1} \frac{2e^{-n \beta m}[1+2\cos(n\theta)]}{n
    |1-e^{in\theta}|^2}.
\end{split}
\nonumber
\end{equation}
Using now $\hat{g}^{\: \mu \mu'} \hat{g}^{\: \nu \nu'} 
\der_{\mu}\der_{\nu'}\sigma 
\der_{\nu}\der_{\mu'}\sigma=1+2\cos(2n\theta)$, 
\begin{equation}
% \label{Delta2}
\begin{split}
  -\ln \det\left(\Delta^{(2)}-m^2\right)&=\int^{\infty}_0 \frac{dt}{t}
  \sum_{n\in\mathbb{Z}}\int d^3x \sqrt{g} \ \hat{g}^{\:
    \mu \mu'} \hat{g}^{\: \nu \nu'} K_{\mu \nu, \mu' \nu'}(t,\sigma(x,\gamma^n x))\\
  &=\frac{5 m^3}{6\pi}{\rm Vol}({\mathbb R^3}/{\mathbb Z})+\sum^{\infty}_{n=1}
  \frac{2e^{-n \beta m}[1+2\cos(n\theta)+2\cos(2n\theta)]}{n
    |1-e^{in\theta}|^2}.
\end{split}
\nonumber
\end{equation}
Putting everything together according to \eqref{putTogether}, the one-loop correction to the classical 
Euclidean action is
\begin{equation}
% \label{1loopg2}
S^{(1)}=\sum^{\infty}_{n=1} \frac{2e^{-n \beta m}[\cos(2n\theta)-\cos(n\theta)]}{n |1-e^{in\theta}|^2}\\
\stackrel{m= 0}{\longrightarrow}  \sum^{\infty}_{n=1} 
\frac{1}{n}\left(\frac{e^{2in\theta}}{1-e^{in\theta}}+\frac{e^{-2in\theta}}{1-e^{-in\theta}}\right).
\nonumber
\end{equation}
The two divergent sums can be made convergent by replacing $\theta$ by
$\theta +i\epsilon$ in the first sum, and by $\theta-i\epsilon$ in the
second one.\footnote{This analytic continuation is quite natural considering the fact that angular potentials 
which are real in Lorentzian signature become imaginary in Euclidean signature.  The partition 
function $Z[\beta, \theta]$ is most naturally viewed as a function of complex angular potential.}
Defining $q=e^{i(\theta+i\epsilon)}$  gives 
\begin{equation}
%   \label{eq:7}
  S^{(1)}=\sum^{\infty}_{n=1} 
\frac{1}{n}\left(\frac{q^{2n}}{1-q^n}+\frac{\bar q^{2n}}{1-\bar
    q^n}\right), 
\nonumber
\end{equation}
or equivalently,
\begin{equation}
\begin{split}
% \label{1loopg2reg}
e^{S^{(1)}}=\prod^{\infty}_{k=2} \frac{1}{|1-q^k|^2}.
\end{split}
\nonumber
\end{equation}
From (\ref{eq:3}) we can then read off the 1-loop partition function around flat space:
\begin{equation}
% \label{1loopg2reg}
 Z[\beta,\theta]=e^{-\frac{1}{\hbar}S^{(0)}+S^{(1)}+O(\hbar)}=e^{\frac{\beta}{8G\hbar}}\prod^{\infty}_{k=2} 
\frac{1}{|1-q^k |^2}\left(1+O(\hbar)\right).
\nonumber
\end{equation}
This expression matches the
vacuum BMS$_3$ character, as computed in \cite{Oblak:2015sea}, for a Euclidean time translation by $\beta$ 
and central charge $c_2=3/G$. It can also be obtained as the flat limit of the one-loop partition function in 
AdS$_3$ \cite{Giombi:2008vd}.

\section*{Acknowledgements}
%\label{sec:acknowledgements}

\addcontentsline{toc}{section}{Acknowledgments}

We thank A. Lepage-Jutier, G. Moore, P. Salgado-Rebolledo and A. Strominger for useful conversations. This 
work is supported in part by the Fund for Scientific
Research-FNRS (Belgium), by IISN-Belgium, by ``Communaut\'e fran\c
caise de Belgique - Actions de Recherche Concert\'ees'' and by the National Science and Engineering Council 
of Canada.

%\bibliography{/Users/gbarnich/Dropbox/Literature/master}
%\bibliography{C:/Users/Glenn/Dropbox/Literature/master}

%\end{document}

\section*{References}
\addcontentsline{toc}{section}{References}

\renewcommand{\section}[2]{}%

\def\cprime{$'$}
\providecommand{\href}[2]{#2}\begingroup\raggedright\endgroup

\end{document}